\documentclass[aps,prl,twocolumn,showpacs,superscriptaddress]{revtex4}

\usepackage {graphicx}
\usepackage {tabularx}
\usepackage{epsf}
\usepackage{braket}
\usepackage{array}

\begin{document}

\title{Electron shakeoff following the $\beta^+$ decay of trapped $^{35}$Ar$^+$  ions}

\newcommand{\lpc}{LPC Caen, ENSICAEN, Universit\'e de Caen,
CNRS/IN2P3, Caen, France}
\newcommand{\celia}{CELIA, Univ. Bordeaux - CNRS UMR 5107 - CEA, F-33405 Talence,  France}
\newcommand{\cimap}{CIMAP, CEA/CNRS/ENSICAEN, BP 5133,
F-14070, Caen, France}
\newcommand{\ganil}{GANIL, CEA/DSM-CNRS/IN2P3, Caen, France}
\newcommand{\granada}{Departamento de F\'{i}sica At\'{o}mica, Molecular y Nuclear,
Universitad de Granada, 1871 Granada, Spain}
\newcommand{\msu}{NSCL and Department of Physics and Astronomy,
Michigan State University, East-Lansing, MI, USA}
\newcommand{\leuven}{Instituut voor Kern- en Stralingsfysica, Katholieke Universiteit Leuven, B-3001 Leuven, Belgium}
\newcommand{\riken}{Atomic Physics laboratory, RIKEN, Saitama 351-0198, Japan}
\author{C.~Couratin}
\author{X.~Fabian}
\affiliation{\lpc}
\author{B. Fabre}
\author{B.~Pons}
\affiliation{\celia}
\author{X.~Fl\'echard}
\email{flechard@lpccaen.in2p3.fr}
\author{E.~Li\'enard}
\author{G.~Ban}
\affiliation{\lpc}
\author{M.~Breitenfeldt}
\affiliation{\leuven}
\author{P.~Delahaye}
\affiliation{\ganil}
\author{D.~Durand}
\affiliation{\lpc}
\author{A.~M\'ery}
\affiliation{\cimap}
\author{O.~Naviliat-Cuncic}
\affiliation{\lpc}
\affiliation{\msu}
\author{T.~Porobic}
\affiliation{\leuven}
\author{G.~Qu\'em\'ener}
\affiliation{\lpc}
\author{D. Rodr\'{i}guez}
\affiliation{\granada}
\author{N.~Severijns}
\affiliation{\leuven}
\author{J-C.~Thomas}
\affiliation{\ganil}
\author{S.~Van Gorp}
\affiliation{\riken}
\date{\today}

\begin{abstract}
The electron shakeoff of $^{35}$Cl atoms resulting
from the $\beta$$^+$ decay of $^{35}$Ar$^+$ ions has been investigated using a Paul trap coupled to a recoil-ion spectrometer. The charge-state
distribution of the recoiling daughter nuclei is compared to theoretical calculations accounting for shakeoff and Auger processes.
The calculations are in excellent agreement with the experimental results and enable to identify the ionization reaction 
routes leading to the formation of all charge states. 
\end{abstract}

\pacs{32.80.Aa, 32.80.Hd, 79.20.Fv}

\maketitle

Precision measurements of the recoil-ion energy spectra in nuclear $\beta$ decay constituted sensitive tools to establish 
the vector  axial vector structure of
the weak interaction \cite{Hamilton47,Allen59,Johnson63}. In particular, these measurements give access to the so called $\beta-\nu$ angular correlation coefficient, $a_{ \beta\nu}$, which is sensitive to scalar and tensor exotic couplings excluded by the Standard Model of elementary particles \cite{Severijns11}. The search for such exotic interactions has motivated new experiments 
using modern trapping techniques coupled to intense radioactive beams with high production rates \cite{Severijns11}. 
Most of ongoing experiments detect the $\beta$ particles and the recoil-ions in coincidence, providing a precise recoil-ion energy measurement using time of flight (TOF) techniques, and a precise control of systematic effects \cite{Gorelov05,Vetter08,Flechard11}.
The LPCTrap setup \cite{Rodriguez06,Flechard08}, installed at GANIL, is based on the use of a Paul trap, to confine radioactive ions, coupled to a recoil-ion spectrometer. It has been  recently upgraded to
perform simultaneous measurements of both the charge-state and the energy of the recoil-ions. Fundamental atomic processes such as electron shakeoff (SO) resulting from the sudden change of the central potential can thus also be addressed through a measurement of the 
charge-state distribution of the recoiling ions.
The setup has already enabled the measurement of electron SO in the decay of $^{6}$He$^+$ \cite{Couratin12}. 
For this ideal textbook case, with only one electron, simple quantum calculations 
based on the sudden approximation (SA) could be tested with a relative precision better than $4{\times} 10^{-4}$. 
Beyond the prototypical $^{6}$He$^+$ case, heavier systems such as $^{35}$Ar$^+$ can reveal the role of more subtle shakeoff dynamics involving 
several electrons, and of subsequent relaxation processes such as the emission of Auger electrons. 
These multi-electron processes, of paramount importance in atomic and molecular physics, have mainly been studied as post-collision mechanisms following 
the absorption of a photon in the X-ray spectral range (see for instance \cite{Carlson66,Saito94,Persoon01,Kaneyasu08}). Alternatively,
SO and Auger processes can conveniently be decoupled within multi-electron rearrangements induced by sudden nuclear $\beta$-decay. 
These ionization
mechanisms have been explored for a large variety of $\beta^-$ emitters \cite{Carlson63}.  In contrast, information is 
scarse for $\beta^+$ decaying parent atoms \cite{Gorelov00,Scielzo03,Vetter08} and even totally missing, up to our knowledge, for multi-electronic singly-charged
systems. We thus investigate here the charge-state distribution following $\beta^+$ emission of $^{35}$Ar$^+$. In addition to be interesting {\em per se}, 
it is worth
noting that SO can further be a source of systematic error for $\beta-\nu$ angular correlation coefficient measurements. This systematic error,
that was found to be small in a previous LPCTrap experiment on $^6$He$^+$ decay \cite{Flechard11}, would become problematic for many electron systems. 
For the WITCH setup \cite{Beck11} installed at ISOLDE-CERN, whose main goal is to measure $a_{ \beta\nu}$ in the decay of $^{35}$Ar$^+$ ions confined in a Penning trap, an independent measurement of the charge-state distribution of the recoil ions will ease the analysis and improve the precision \cite{VanGorp13}. 
\begin{figure}[!ht]
\includegraphics[width=85mm]{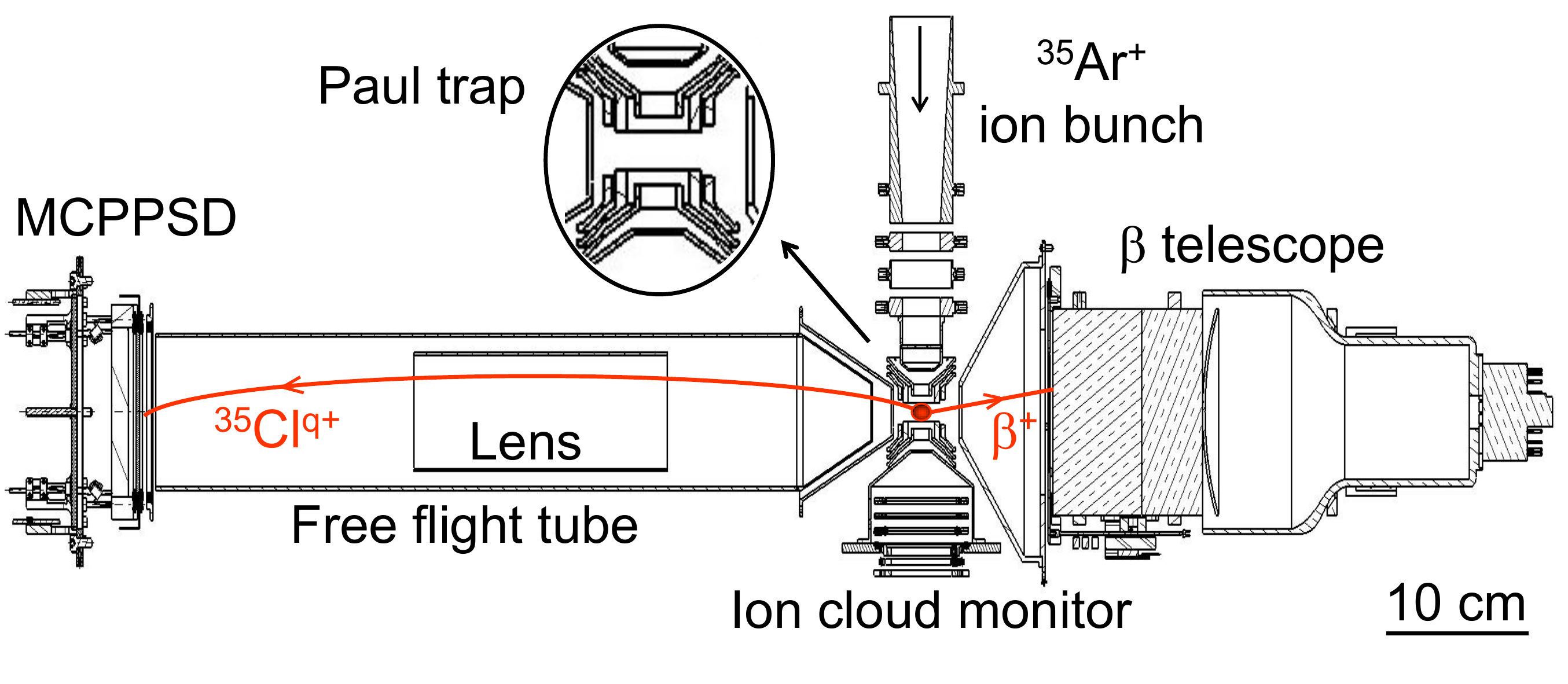}
\caption{(Colour online). Top view of the experimental setup. The insert shows the structure
of the six stainless steel rings of the Paul trap. See text for details.}
\label{fig.1}
\end{figure}

The experimental setup has been described
in detail previously \cite{Couratin12,Flechard08,Rodriguez06}.  The radioactive $^{35}$Ar nuclei were produced
at the SPIRAL target-ECR ion source system of GANIL, Caen, France, by a primary beam of  $^{36}$Ar ions impinging on a graphite target. After
mass separation the $^{35}$Ar$^+$ ions were guided at 10 keV
through the LIRAT low energy beam line. At the entrance of the LPCTrap
apparatus, the $^{35}$Ar$^+$ beam intensity was typically  $10^7$~pps. The ions were first injected in a Radio
Frequency Cooler and Buncher (RFQCB) \cite{Darius04} for the beam preparation.
This linear Paul trap, mounted on a high voltage platform to
decelerate the ions down to 50 eV, was filled with He buffer gas
at a pressure of $1.6\times10^{-2}$ mbar
to cool down the ions below 1 eV.
The $^{35}$Ar$^+$ ions from LIRAT were continuously injected in the RFQCB,
cooled, and accumulated into bunches before being
extracted with a cycle period of 200~ms. They were then reaccelerated downstream
using a pulsed cavity, transported between the two traps with a kinetic energy of about 1~keV, and decelerated down to
100~eV by a second pulsed cavity located at the entrance of the measurement transparent Paul trap (MTPT).
For each injection cycle, an average of about $2 \times 10^3$ $^{35}$Ar$^+$ ions were
successfully trapped and
confined by applying a 0.48 MHz RF voltage of 120 V$_{pp}$ to the two inner rings of the MTPT (Fig. \ref{fig.1}).
Helium buffer gas at a pressure of $10^{-5}$ mbar was
also used in the MTPT chamber
to further cool down the trapped ions.
The $\beta$ particles and the recoiling ions resulting from the $\beta$ decay
of the trapped $^{35}$Ar$^+$ ions were detected in coincidence using detectors located
around the trap (Fig.\ref{fig.1}). The $\beta$ telescope, composed of a thin
double sided silicon strip detector followed by a plastic scintillator,
provides the position and the energy of the incoming $\beta$ particles. The signal
from the plastic scintillator also defines the
reference time for a decay event. A recoil ion spectrometer enables to separate the
charge-states of the recoiling ions using their time of flight (TOF). Ions emitted
towards the recoil ion spectrometer cross a first collimator through a 90\% transmission
grid (set at ground potential). They are then accelerated by a $-2$~kV potential
applied to a free flight tube (Fig. \ref{fig.1}) whose entrance and exit are defined by 
two additional 90\% transmission grids.  Inside the tube, an electrostatic lens
at $-250$~V permits a 100\% collection efficiency of the ions by
a micro-channel plate position sensitive detector (MCPPSD) \cite{Lienard05}.
A $-4$~kV voltage applied on the
front plate of the MCPPSD ensures a detection efficiency close to maximum for all
charge-states of the recoil ions, independently of their initial kinetic energy. 
\begin{figure}[!ht]
\includegraphics[width=85mm]{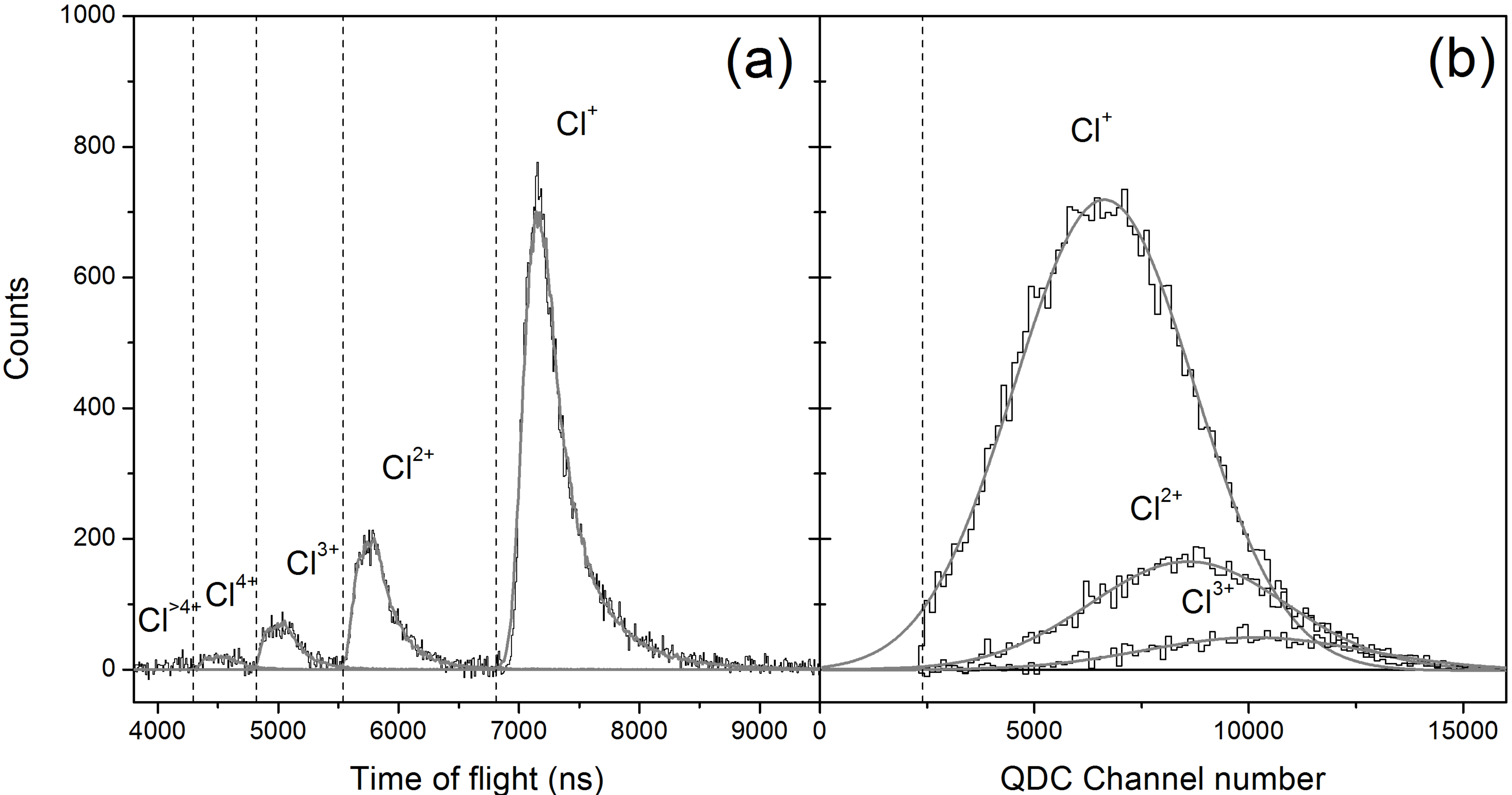}
\caption{(a) Experimental (black line)  and simulated TOF spectra associated to the different charge-states (gray lines); 
vertical dashed lines indicate the ranges of integration used to obtain the charge-state branching ratios. (b) 
Charge collected from the  MCPPSD for different recoil ion charge-states (black lines)  adjusted with gaussians (gray lines); 
the vertical dashed line indicates the cut due to electronic threshold.}
\label{fig.2}
\end{figure}

For each
detected event, the energy and position of the $\beta$ particle, the time of flight (TOF) and position
of the recoil ion, were recorded. 
The procedure applied for the detector calibrations was identical to that described in Ref. \cite{Flechard11}. Only events corresponding
to a $\beta$ particle depositing more than 0.4~MeV in the scintillator were kept in the analysis. 
The TOF distribution measured for the $^{35}$Cl$^{q+}$ recoil ions resulting from  $^{35}$Ar$^+$ 
$\beta$  decay is shown in Fig. \ref{fig.2}(a). A constant background in TOF due to uncorrelated signals from the recoil ion and $\beta$
detectors has been subtracted from the data.
%
%

The experimental charge-state branching ratios and their associated statistical uncertainty were simply deduced from the integration of counts
within the TOF selection windows displayed on Fig.\ref{fig.2}(a). Two additional corrections, labeled Tail$_{corr.}$ and MCP$_{corr.}$ in Table I, were then applied. The first takes into account the tails of charge distributions extending beyond
their respective integration windows. TOF spectra associated to each charge-state were generated using Monte-Carlo
simulations~\cite{Flechard11,Couratin12} and were adjusted to the experimental data. Several ingredients of the simulations, such as the exact size of the trapped  ion cloud, weak decay
branches of  $^{35}$Ar towards excited states, and possible scattering of the $\beta$ particles on parts of the trapping chamber were neglected or approximated. Therefore, a conservative relative uncertainty of 10~\% was applied on these corrections.
The second correction concerns the dependence of the MCPPSD detection efficiency on the charge-state of the recoil ions. 
The loss of detected events due to electronic threshold was precisely estimated by fitting the charge distributions collected from the recoil ion detector with gaussian functions (Fig. \ref{fig.2}(b)). The data were then corrected for the relative detection efficiency obtained for each charge-state.
For charged recoil ions, the experimental charge-state branching ratios including these corrections are given in Table I. Electron capture probabilities from He buffer gas between the Paul trap and the spectrometer have also been estimated using experimental cross sections measured for Kr$^{q+}$ + He collisions in the same velocity regime~\cite{Justiniano84}.
Kr and Cl having similar ionization potentials, the charge exchange cross sections for Kr$^{q+}$ ions constitute a good approximation of what one would expect with Cl$^{q+}$ ions.
Even for the higher charge states involved here (q=5 and 6), these probabilities are only of the order of a few 10$^{-4}$. They were therefore  neglected at the present level of precision.
For a dominant part of the decay events, there is no electron shakeoff and the $\beta$$^+$ decay of a $^{35}$Ar$^+$ ion results in the recoil of a neutral $^{35}$Cl atom.
For Cl recoil atoms, both the collection efficiency and the MCPPSD detection efficiency are very low, and depends strongly on their initial energy. 
The TOF associated to such detected events being larger than 11$\mu$s, they do not appear on the spectrum shown in the Fig. \ref{fig.2}(a).
To estimate the number of $^{35}$Cl atoms produced during the experiment, we have used the number of $\beta$ particles detected in "singles" (without condition on the detection of a recoil).
Knowing the overall absolute detection efficiency for ions, the fraction of  the "singles" events associated to charged $^{35}$Cl recoils could be inferred, the rest being associated to $^{35}$Cl atoms.
This estimate leads to 72(10)\% of neutral $^{35}$Cl recoils, with an uncertainty dominated here by the error on the overall ion detection efficiency. 
This result is in good agreement with the 73.9\% ratio obtained from the theoretical calculations that we detail hereinafter. 
{\setlength{\tabcolsep}{0.8em}
\begin{table*}
\caption{Experimental ion charge-state relative branching ratios (\%) and included corrections (see text) compared to calculations with and without recoil and Auger ionizations.}
\begin{tabularx}{\textwidth}{cccccccc}
\hline
\hline
Charge  & MCP$_{corr.}$ & Tail$_{corr.}$ &  Exp. &  With recoil & Without recoil & With recoil & Without recoil\\
 & & & results & With Auger & With Auger & Without Auger & Without Auger\\
\hline
1 &  0.37 & -0.17 & 74.75 $\pm$1.07 & 74.37 & 74.44 & 87.07 & 87.37 \\
2 &  -0.24 & -0.09 & 17.24  $\pm$0.44 & 16.98 & 16.91 & 11.92 & 11.66 \\
3 &  -0.09 & 0.03 & 5.71 $\pm$0.27 & 6.03 & 6.04 & 0.95 & 0.91 \\
4 &  -0.03 & 0.13 & 1.58 $\pm$0.21 & 1.79 & 1.79 & 0.05 & 0.05 \\
$>$4 &  -0.01 & 0.10 & 0.71 $\pm$0.18 & 0.82 & 0.82 & $<$0.002 & $<$0.002 \\
\hline
\hline
\end{tabularx}
\end{table*}

Subsequently to the sudden decay of $^{35}$Ar$^+$, primary Cl$^{q+}$ ions are formed by ionization. In the framework of the independent electron 
model (IPM,\cite{Bransden}), well suited to describe the dynamics of multielectronic systems, the probability to ionize $q_S=q$ electrons
among the $N=17$ total ones reads
\begin{equation}
P_{q_S}^{ion}=\sum_{i_1=1}^{N}p_{i_1} \sum_{i_2>i_1}^{N}p_{i_2} \dots \sum_{i_{q_S}>i_{q_S-1}}^{N}p_{i_{q_S}} \prod_{j\neq i_1,\dots,i_{q_S}}^{N}{\left ( 1 - p_{j} \right )}
\label{pSO}
\end{equation}
where $p_i$ is the one-electron ionization probability for the $i^{th}$ electron. In our work, $p_i$ results from SO, with the 
underlying assumptions: (i) the so-called direct ionization mechanism, in which the $\beta$ particle knocks out orbital electrons, is neglected, 
and (ii) shakeup processes, which would imply electron excitation(s) as a result of the $\beta$-decay, are also neglected. Assumption (i) is consistent with 
the fact that the $\beta$ emission energy (with end point $E_\beta^{max}=4.94$ MeV) is considerably larger than the energy of bound electrons,
so that direct ionization is unlikely \cite{Freedman74}. 
Most of the inelastic processes involving electron vacancies in intermediate- and large-$Z$ species consist of transitions to the continuum \cite{Carlson63bis},
justifying (ii). Therefore, the one-electron ionization probability $p_i$, with $i$ initially belonging to the ($n_i,l_i$) subshell, is expressed as
\begin{equation}
p_i=1- \sum_{n'\le 3}{| \langle \varphi^{(Cl)}_{n'l} | \text{e}^{-\text{i} {\bf K.r}} |  \varphi^{(Ar^+)}_{n_il_i} \rangle |^2}
\label{pi}
\end{equation}
in the rest frame of the daughter nucleus of mass $M$ which recoils with energy $E_R$ and momentum $K=\sqrt{2E_R/M}$ (in atomic units). 
$\varphi^{(Ar^+,Cl)}_{nl}$ is the wavefunction describing one electron orbiting in the $nl$-subshell of Ar$^+$ or Cl. Because of the small
values of $K$ ($E_R^{max}=452$ eV), $\text{e}^{-\text{i} {\bf K.r}}$ can be expanded in eq. (\ref{pi}) in order to highlight the 
mechanisms underlying $\beta$-induced SO. Up to second order in $K^2$, we obtain \cite{carlson63recoil}
{\setlength\arraycolsep{2pt}
\begin{eqnarray}
p_i&=&1- \sum_{n'\le 3} \Big\{ | \langle \varphi^{(Cl)}_{n'l_i} | \varphi^{(Ar^+)}_{n_il_i} \rangle |^2 \nonumber \\
&+&K^2 | \langle \varphi^{(Cl)}_{n'l_i \pm 1} | {\bf r} |  \varphi^{(Ar^+)}_{n_il_i} \rangle |^2 \\
&-&K^2 \text{Re}\langle\varphi^{(Cl)}_{n'l_i}|\varphi^{(Ar^+)}_{n_il_i}\rangle^\ast \langle \varphi^{(Cl)}_{n'l_i} | r^2 |  \varphi^{(Ar^+)}_{n_il_i} \rangle \Big\} \nonumber.  
\label{dev}
\end{eqnarray}}
It is thus clear that ionization stems from the coherent superposition of two effects: the static Ar$^+$/Cl orbital mismatch, through the 
$| \langle \varphi^{(Cl)}_{n'l_i} | \varphi^{(Ar^+)}_{n_il_i} \rangle |^2$ terms, and the recoil of the Cl daughter nucleus, through 
the $K^2$-dependent terms in (3).

For the calculation of the recoil-induced ionization terms involved in (3), we have used the mean recoil energy 
obtained in our experiment, $\bar{E}_R=376$ eV, to define the numerical value of $K$ (0.02074 a.u.). As Ar$^+$ and Cl are open-shell
valence systems, the wavefunctions $\varphi^{(Ar^+,Cl)}_{nl}$ have been computed by means of Restricted Open-shell Hartree-Fock (ROHF) 
calculations using the GAMESS-US quantum-chemistry package \cite{Gamess}. These wavefunctions can alternatively be obtained in terms of 
Unrestricted Hartree-Fock (UHF) computations \cite{pople54}, and we found that the ROHF and UHF $p_i$ probabilities differ by less than 1\% for all the $1s,...,3p$ levels \cite{note_clementi}.  
%
%

Once inner-shell vacancies have been created in Cl through ionization, radiative and Auger transitions involving higher-lying electrons tend to fill these
vacancies. The probabilities associated to these transitions have to be properly introduced in our calculations, especially as Auger processes are known to
contribute significantly to the production of high charge states \cite{Scielzo03}. 
Kaastra and Mewe \cite{Kaastra93} have computed the probabilities $\tilde{p}^{s,m_i}_i$ corresponding to the ejection of $m_i$ electrons through Auger cascades after the 
electron $i$, element of the inner $n_il_i$-subshell, has been removed from Cl$^{s+}$. $s$ is the ionization stage of Cl and corresponds to the number of electrons 
previously pulled out from the outermost subshells. 
Consistently with the treatment of multiple SO ionization, the description of multiple vacancies, and related Auger cascades, is performed in the IPM framework. 
The probability for ejection of $q_S$ electrons by SO followed by Auger removal of $q_A$ electrons thus reads
\begin{eqnarray}
P_{q_S,q_A} & = & \sum_{m_{i_1},\dots,m_{i_{q_S}} \atop m_{i_1}+\dots+m_{i_{q_S}}=q_A}\sum_{i_1=1}^{N}{p_{i_1} \tilde{p}_{i_1}^{s,m_{i_1}}} \sum_{i_2>i_1}^{N}{p_{i_2} \tilde{p}_{i_2}^{s,m_{i_2}}} \dots  \nonumber \\
&  & \sum_{i_{q_S}>i_{q_S-1}}^{N}p_{i_{q_S}} \tilde{p}_{i_{q_S}}^{s,m_{i_{q_S}}} \prod_{j\neq i_1,\dots,i_{q_S}}^{N}{\left ( 1 - p_j \right )}
\label{pfinal}
\end{eqnarray}
where $q_S+q_A=q$, the charge state finally observed. 

The computed charge-state branching ratios, which consist of the relative populations of Cl$^{q+}$ species 
among the total ion yield, are compared to their experimental counterparts in Table I. Eq. (\ref{pfinal}) provides a very good agreement with
the measurements. 

{\setlength{\tabcolsep}{0.5em}
\begin{table*}
\caption{Main ionization routes leading to Cl$^{q+}$ formation (in \%). $nl^{-1}$ refers to primary SO hole creation in the $nl$-subshell of Cl 
while $m \times e_A$ means emission of $m$ Auger electrons. The asterisk indicates multiple Auger emission.}
\begin{tabularx}{\textwidth}{ccccc}
\hline
\hline
Cl$^+$ & Cl$^{2+}$ & Cl$^{3+}$ & Cl$^{4+}$ & Cl$^{5+}$\\
\hline
$3p^{-1}$ : \textbf{79.83}  & $2p^{-1}+1e_A$ : \textbf{49.71} & $2s^{-1}+2e^*_A$ : \textbf{52.21} & $2s^{-1}3p^{-1}+2e^*_A$ : \textbf{36.95} & $1s^{-1}+4e^*_A$ : \textbf{35.56} \\
$3s^{-1}$ : \textbf{20.16} & $3p^{-2}$ : \textbf{29.32} &$ 2p^{-1}3p^{-1}+1e_A$ : \textbf{29.36} & $1s^{-1}+3e^*_A$ : \textbf{31.02} & $1s^{-1}3p^{-1}+3e^*_A$ : \textbf{19.69}\\
 & $3s^{-1}3p^{-1}$ : \textbf{18.51} & $2p^{-1}3s^{-1}+1e_A$ : \textbf{7.41} & $2s^{-1}3s^{-1}+2e^*_A$ : \textbf{9.31} & $2s^{-1}2p^{-1}+3e^*_A$ : \textbf{15.64} \\
 & $3s^{-2}$ : \textbf{1.17} & $3s^{-1}3p^{-2}$ : \textbf{4.37} & $2p^{-1}3p^{-2}+1e_A$ : \textbf{8.28} & $2s^{-1}3p^{-2}+2e^*_A$ : \textbf{9.26} \\
 & & $3p^{-3}$ : \textbf{3.46} & $2p^{-2}+2e_A$ : \textbf{5.86} & $2s^{-1}3s^{-1}3p^{-1}+2e^*_A$ : \textbf{5.84} \\
 & & $1s^{-1}+2e^*_A$ : \textbf{1.56} & $2p^{-1}3s^{-1}3p^{-1}+1e_A$ : \textbf{5.23} & $1s^{-1}3s^{-1}+3e^*_A$ : \textbf{4.90}\\
 & & & & $2p^{-2}3p^{-1}+2e^*_A$ : \textbf{3.66} \\
 & & & & $2p^{-1}3s^{-1}3p^{-2}+1e_A$ : \textbf{1.31} \\
\hline
\hline
\end{tabularx}
\end{table*}
 
The calculations can then be used to discrimate between the roles of SO and Auger transitions in the production of Cl$^{q+}$ ions. Neglecting Auger
decays, {\em i.e.} computing the charge-state distribution according to Eq. (\ref{pSO}), severely distorts the Cl$^{q+}$ populations: the $q=+1$ population
is overestimated by $\sim 13 \%$ while multiple SO is inefficient to explain the abundance of $q \ge 3$ states. Such an importance of Auger processes has to be contrasted 
with previous studies on lighter systems where differences between full and SO-restricted calculations did not exceed a few percent (see e.g. \cite{Scielzo03}). 
The increase of the Auger importance with increasing $Z$ can simply be related to the higher multiplicity of Auger cascades; for instance, we derive from
\cite{Kaastra93} that Cl$^+(1s^{-1})$ preferentially stabilizes by emitting 3 electrons while Na$^+(1s^{-1})$ relaxes by ejecting only 1 electron. 

The calculations can also be employed to estimate the role of recoil-induced ionization by artificially setting $K=0$ in eq. (\ref{dev}). 
The static orbital mismatch explains most of SO ionization (see Table I). However, accounting for the recoil changes the final population of neutral Cl from
74.5 to 73.9$\%$. Even if it looks small at first sight, such a variation can significantly influence the precise determination of the $a_{\beta\nu}$ 
correlation coefficient \cite{Scielzo03}. The nuclear recoil is almost inconsequential to the charge-state distribution (Table I). Multiple 
ionization from primary SO is small ($\sim 3\%$) so that the relative populations of Cl$^{q+}$ high charge states are mainly monitored by the 
subsequent Auger cascades. 
 
Finally, we can search within Eq. (\ref{pfinal}) the electronic probabilities which mostly contribute
to the formation of a given Cl$^{q+}$ state. The ionization routes contributing more than 1\% to the formation of Cl$^{q+}$, with $1 \le q \le 5$, are presented
in Table II. Single SO from $n=3$ states explains $\sim$100$\%$ of Cl$^+$ formation since ionization in 
inner shells leads to the formation of higher charge states through Auger cascades with almost $100\%$ probability. For Cl$^{2+}$, 
twofold SO from the outer $n=3$ shell, 
eventually followed by radiative stabilization when $3s$ holes are involved, represents $\sim$50$\%$ of the population and single Auger transitions filling 
the $2p^{-1}$ SO-hole make the rest. The relevance of SO ionization with multiplicity greater than 2 rapidly falls down (see Table I). 
As a result, the creation of Cl$^{q+}$ ions with $q \ge 3$ mostly involves Auger decays subsequent to one- and two-fold SO ionization. Moreover, 
multiple Auger emission, involving intermediate core-hole states and emission of several electrons during a single-hole decay, becomes increasingly important 
for high $q$; it participates to almost 100\% of Cl$^{5+}$ ion creation. 

To sum up, our joined experimental/theoretical endeavour has provided a quite complete picture of ion formation resulting from the
$\beta^+$ decay of $^{35}$Ar$^+$. We plan to apply the same techniques to $^{19}$Ne$^+$ and also revisit previous 
studies \cite{Scielzo03} in the near future. Besides the intrinsic interest of such investigations to nuclear physics, this will allow obtaining 
a more complete and $Z$-dependent picture of the underlying ionization mechanisms. 

\begin{acknowledgments}
The authors thank the LPC staff for their strong involvement in the
LPCTrap project and the GANIL staff for the preparation of a high quality ion beam.
The authors also acknowledge the computational facilities provided by the M\'esocentre de Calcul Intensif Aquitain at 
University of Bordeaux \cite{mcia}. 
D.R. acknowledges support from the Spanish ministry of Economy and competitiveness under the project
FPA2010-14803 and the action AIC10-D-000562.
\end{acknowledgments}

\begin{thebibliography}{0}
\expandafter\ifx\csname natexlab\endcsname\relax\def\natexlab#1{#1}\fi
\expandafter\ifx\csname bibnamefont\endcsname\relax
  \def\bibnamefont#1{#1}\fi
\expandafter\ifx\csname bibfnamefont\endcsname\relax
  \def\bibfnamefont#1{#1}\fi
\expandafter\ifx\csname citenamefont\endcsname\relax
  \def\citenamefont#1{#1}\fi
\expandafter\ifx\csname url\endcsname\relax
  \def\url#1{\texttt{#1}}\fi
\expandafter\ifx\csname urlprefix\endcsname\relax\def\urlprefix{URL }\fi
\providecommand{\bibinfo}[2]{#2}
\providecommand{\eprint}[2][]{\url{#2}}

\end{thebibliography}


\begin{thebibliography}{999}
%
\bibitem{Hamilton47} D. R. Hamilton, Phys. Rev. {\bf 71}, 456 (1947).
\bibitem{Allen59} J. S. Allen {\em et al}, Phys. Rev. {\bf 116}, 134 (1959).
\bibitem{Johnson63} C.H. Johnson {\em et al.}, Phys. Rev. {\bf132}, 1149 (1963).
\bibitem{Severijns11} N. Severijns and O. Naviliat-Cuncic, Annu. Rev. Nucl. Part. Sci {\bf 61}, 23 (2011).
\bibitem{Gorelov05} A. Gorelov {\em et al.}, Phys. Rev. Lett. {\bf 94}, 142501 (2005).
\bibitem{Vetter08} P.A. Vetter {\em et al.}, Phys. Rev. C {\bf 77}, 035502 (2008).
\bibitem{Flechard11}X. Fl\'echard {\em et al.}, J. Phys. G {\bf 38}, 055101 (2011).
\bibitem{Rodriguez06} D. Rodr\'{i}guez {\it et al.}, Nucl. Instrum. Methods Phys. Res. A {\bf 565}, 876 (2006).
\bibitem{Flechard08} X. Fl\'echard {\it et al.}, Phys. Rev. Lett. {\bf 101}, 212504 (2008).
\bibitem{Couratin12} C. Couratin {\em et al.}, Phys. Rev. Lett. {\bf 108}, 243201 (2012).
\bibitem{Carlson66} T. A. Carlson, W. E. Hunt and M. O. Krause, Phys. Rev. {\bf 151}, 41 (1966).
\bibitem{Saito94} N. Saito and I. H. Suzuki, Physica Scripta {\bf 49}, 80 (1994).
\bibitem{Persoon01} P. Persoon {\em et al.}, Protein Science {\bf 10}, 2480 (2001).
\bibitem{Kaneyasu08} T. Kaneyasu {\em et al.}, J. Phys. B {\bf 41}, 135101 (2008).
\bibitem{Carlson63} T. A. Carlson, Phys. Rev. {\bf 131}, 676 (1963) {\em and references therein}.
\bibitem{Gorelov00} A. Gorelov {\em et al.}, Hyperfine Interact. {\bf 127}, 373 (2000).
\bibitem{Scielzo03} N. D. Scielzo {\em et al.}, Phys. Rev. A {\bf 68}, 022716 (2003).
\bibitem{Beck11} M. Beck {\em et al.}, Eur. Phys. J. A {\bf 47}, 45 (2011).
\bibitem{VanGorp13} S. Van Gorp {\em et al.}, Phys. Rev. Lett. (2013, submitted).
\bibitem{Darius04} G. Darius {\it et al.}, Rev. Sci. Instrum. {\bf 75}, 4804 (2004).
\bibitem{Lienard05} E. Li\'enard {\it et al.}, Nucl. Instrum. Methods Phys. Res. A {\bf 551}, 375 (2005).
\bibitem{Justiniano84}  E. Justiniano {\it et al.}, Phys. Rev. A {\bf 29}, 1088 (1984).
\bibitem{Bransden} B. H. Bransden and M. H. C. McDowell, {\em Charge Exchange and the Theory of Ion-Atom Collisions} (Clarendon, Oxford, 1992).
\bibitem{Freedman74} M. S. Freedman, Annual Review of Nuclear Science {\bf 24}, 209 (1974). 
\bibitem{Carlson63bis} T. A. Carlson, Phys. Rev. {\bf 130}, 2361 (1963).
\bibitem{carlson63recoil} T. A. Carlson, F. Pleasonton and C. H. Johnson, Phys. Rev. {\bf 129}, 2220 (1963).
\bibitem{Gamess} M. W. Schmidt {\em et al.}, J. Comput. Chem. {\bf 14}, 1347 (1993).
\bibitem{pople54} J. A. Pople and R. K. Nesbet, J. Chem. Phys. {\bf 22}, 571 (1954).
\bibitem{note_clementi} It is quite common to employ the Roothan-Hartree-Fock (RHF) orbitals of Clementi and Roetti \cite{clementi74} to compute the ionization probabilities $p_i$. Even though we employ a different (and larger) underlying basis to construct the Ar$^+$ and Cl orbitals, we have verified that the $p_i$ values obtained with the orbitals of Clementi and Roetti differ by less than 1$\%$ from our ROHF and UHF values. 
\bibitem{clementi74} E. Clementi and C. Roetti, At. Data and Nucl. Data Tables {\bf 14}, 177 (1974).
\bibitem{Kaastra93} J. S. Kaastra and R. Mewe, Astron. Astroph. Supp. Ser. {\bf 97}, 443 (1993).
\bibitem{mcia} http://www.mcia.univ-bordeaux.fr
\end{thebibliography}
%

%
\end{document}